# Heterogeneity in Women's Nighttime Ride-Hailing Intention: Evidence from an LC-ICLV Model Analysis

Ke Wang[1], Dongmin Yao[1], Xin Ye[2], Mingyang Pei[3*]

[1]Business School, University of Shanghai for Science and Technology, Shanghai, China,200093
[2]Key Laboratory of Road and Traffic Engineering of Ministry of Education, College of Transportation Engineering, Tongji University, Shanghai, China, 201804
[3]Department of Civil and Transportation Engineering, South China University of Technology, Guangzhou, China, 510641

Emails: kewang053l@foxmail.com, YDominic@163.com, xye@tongji.edu.cn, mingyang@scut.edu.cn

[*]Correspondence: mingyang@scut.edu.cn

**ABSTRACT**

While ride-hailing services offer increased travel flexibility and convenience, persistent nighttime safety concerns significantly reduce women's willingness to use them. Existing research often treats women as a homogeneous group, neglecting the heterogeneity in their decision-making processes. To address this gap, this study develops the Latent Class Integrated Choice and Latent Variable (LC-ICLV) model with a mixed Logit kernel, combined with an ordered Probit model for attitudinal indicators, to capture unobserved heterogeneity in women's nighttime ride-hailing decisions. Based on panel data from 543 respondents across 29 provinces in China, the analysis identifies two distinct female subgroups. The first, labeled the "Attribute-Sensitive Group", consists mainly of young women and students from first- and second-tier cities. Their choices are primarily influenced by observable service attributes such as price and waiting time, but they exhibit reduced usage intention when matched with female drivers, possibly reflecting deeper safety heuristics. The second, the "Perception-Sensitive Group", includes older working women and residents of less urbanized areas. Their decisions are shaped by perceived risk and safety concerns; notably, high-frequency use or essential nighttime commuting needs may reinforce rather than alleviate avoidance behaviors. The findings underscore the need for differentiated strategies: platforms should tailor safety features and user interfaces by subgroup, policymakers must develop targeted interventions, and female users can benefit from more personalized risk mitigation strategies. This study offers empirical evidence to advance gender-responsive mobility policy and improve the inclusivity of ride-hailing services in urban nighttime contexts.

# 1. INTRODUCTION

Ride-hailing services, as a core component of the sharing economy, leverage mobile applications to match drivers and passengers in real time, significantly enhancing the convenience, comfort, and flexibility of urban travel (Shaheen, 2018; Rodier, 2018). Moreover, their innovative service model addresses the traditional "last-mile problem" in public transport (Acheampong et al., 2020), serves as a driving force for global transportation transformation (Tirachini, 2019), and fills public transport gaps in developing countries (Acheampong et al., 2020; Kumar et al., 2022). Despite the operational benefits of ride-hailing, its rapid expansion has been accompanied by growing safety challenges. High-profile incidents of sexual harassment and violent crimes (e.g., Uber's disclosure of nearly 6,000 sexual assault incidents between 2017-2018, criminal incidents involving Didi in China) (Conger, 2019; Grothaus, 2018; Zhang & Munroe, 2018) have raised critical questions regarding platform accountability, driver screening, and trip monitoring (Chaudhry et al., 2018). Women, as a vulnerable group, face heightened risks of crime within the confined space of vehicles (UN Women, 2017; Noor & Iamtrakul, 2023). Survey evidence suggests that 41% of women perceive ride-hailing as "unsafe" (compared to only 25% of men), with fears of sexual assault, abduction, and harassment driving significant nighttime avoidance; 56% of women avoid late-night trips (Young & Farber, 2019; d'Arbois de Jubainville & Vanier, 2017). Such safety anxieties directly suppress usage intention, forcing women to either forego essential mobility or resort to alternative travel options (Plyushteva & Boussauw, 2020; Ceccato & Loukaitou-Sideris, 2022).

This contradiction reveals a core scientific problem: while the inherent advantages of ride-hailing encourage women's adoption, its nighttime safety risks trigger strong behavioral resistance, exhibiting substantial heterogeneity within the female user group. While existing research has examined general ride-hailing behavior, the internal differentiation within the female population remains underexamined. In particular, the heterogeneity mechanisms underlying women's nighttime ride-hailing usage intention are insufficiently quantified. Although studies have identified significant differentiation within the female population based on age, income, geography, and cultural background (Meshram et al., 2020; Ouali et al., 2020; Dunckel-Graglia, 2013), the predominant reliance on methods assuming behavioral homogeneity imposes significant limitations, hindering a deeper investigation into this heterogeneity mechanism. Traditional statistical analyses (e.g., logistic or Poisson regression) typically assume population homogeneity and are thus limited in their ability to identify distinct subgroups within the female population that differ in risk perception and decision-making preferences (Meshram et al., 2020; Mao et al., 2021; Giacomantonio et al., 2024; Liu et al., 2022; Chowdhury & Van Wee, 2020). While latent variable modeling techniques (e.g., factor analysis, structural equation modeling) (Mazzulla et al., 2024;

Wang et al., 2023; Ceccato & Loukaitou-Sideris, 2022; Ouali et al., 2022) have been employed to capture psychological constructs, these attitudinal dimensions are rarely dynamically integrated into behavioral choice models, making it difficult to quantify their causal influence on specific travel decisions. Therefore, there is a critical need for analytical frameworks that can simultaneously uncover unobserved heterogeneity and integrate psychological factors into choice modeling. Such approaches are essential to quantifying the differentiated decision mechanisms of diverse female subgroups regarding nighttime ride-hailing intentions.

To reveal the influence of latent psychological factors on travel behavior, Ben-Akiva et al. (2002) proposed the Integrated Choice and Latent Variable (ICLV) model, which combines structural equation modeling with discrete choice frameworks. This integration allows for a more accurate representation and prediction of behavioral decisions by accounting for unobserved attitudinal factors. The ICLV model framework has been widely applied to travelers' transportation choices and consumers' willingness-to-pay (Gao et.al., 2020; Biancolin et al., 2025; Huang & Qian, 2025; Mohiuddin et al., 2025). Building on this framework, our prior research (Wang et al., 2025) has also conducted a preliminary analysis of the determinants and heterogeneity in women's nighttime ride-hailing intentions. However, two key limitations constrained that study. First, the use of a sequential estimation approach failed to account for measurement error in latent variables, potentially resulting in biased parameter estimates. Second, Likert-scale responses (e.g., 1-5 scores) were treated as continuous rather than ordinal categorical variables, violating their inherent structure and introducing additional estimation bias in the latent constructs. To address these issues, this study employs a Latent Class Integrated Choice and Latent Variable (LC-ICLV) modeling approach. This approach first uses a Latent Class Model (LCM) to identify unobserved subgroups within the female population, revealing distinct decision mechanisms across subgroups. To appropriately handle ordinal indicator variables, the measurement component utilizes ordered Probit models, following Generalized Heterogeneous Data Model (GHDM, Bhat 2015). A full information maximum likelihood estimation is then applied to simultaneously estimate both latent variables and class-specific choice behavior. This integrated approach yields consistent parameter estimates and improves the explanatory and predictive performance of the model in capturing the heterogeneity in women's nighttime ride-hailing choice behavior.

This study makes three key contributions. First, it applies the LC-ICLV framework to systematically uncover heterogeneity in women's nighttime ride-hailing intentions by identifying latent subgroups with distinct risk preferences and decision-making mechanisms using the latent class model. Within each class, mixed logit models capture random taste variation in the influence of key explanatory variables. Second, the measurement component employs ordered Probit models to appropriately handle ordinal attitudinal indicators, and a simultaneous estimation approach is used to ensure consistent and unbiased parameter estimates. This overcomes the biases of traditional methods and achieves unbiased quantification of the differential influence intensity and direction of these latent variables on choice behavior within each subgroup. Third, by revealing class-specific psychological drivers of ride-hailing choices, the findings offer actionable insights for improving women's nighttime travel safety, informing user-centered service design, and guiding targeted policy interventions.

The remainder of this paper is structured as follows. Section 2 reviews literature on heterogeneity in women's travel decision-making. Section 3 describes the data and sample characteristics. Section 4 details the modeling framework. Section 5 presents model estimation results, followed by a discussion and policy implications in Section 6. Section 7 concludes with a summary and research limitations.

## 2．LITERATURE REVIEW

### 2.1 Socioeconomic Characteristics Heterogeneity

Socioeconomic characteristics within the female population profoundly influence their nighttime ride-hailing intentions, with age, income, education level, and occupational status interacting in complex ways. Young women (18-30 years) are primary users of ride-hailing services. Despite higher usage frequency, they exhibit heightened sensitivity to safety risks (Sá & Pitombo, 2019). Compared to women over 40, they report lower perceived safety scores and are more inclined to adopt protective strategies such as traveling with companions (Yang et al., 2022). This heightened concern stems from younger women's stronger social mobility demands and increased exposure to media reports of crime (Yavuz & Welch, 2010). Safety anxieties are particularly acute among low-income women aged 18-24 residing in high-crime neighborhoods (Yates & Ceccato, 2020). Education influences risk response through cognitive processing capacity. Highly educated women (postgraduate and above) place greater emphasis on driver criminal records and platform complaint mechanisms, and are 3.76 times more likely to avoid solo nighttime rides than less-educated groups (Ceccato et al., 2021; Kash, 2019). In contrast, women with lower education levels, often facing information asymmetry and economic constraints, are more likely to accept higher-risk travel (Acheampong et al., 2020). Occupational status also shapes behavioral differences through economic capacity and travel necessity rigidity. Students, while less sensitive to driver appearance, exhibit only 60% of the willingness-to-pay of faculty/staff due to financial limitations (Corazza & D'Eramo, 2025). Employed women demonstrate greater nighttime service acceptance due to commuting need but frequently adjust travel plans to mitigate perceived risks (Wang et al., 2020). Interestingly, family structure significantly shapes decision-making. Single mothers, bound by childcare responsibilities, heavily rely on ride-hailing for urgent nighttime trips (e.g., hospital visits), while women in dual-earner couples may face less travel pressure due to spousal support (Kamp

Dush et al., 2018; Latshaw & Hale, 2016). This socioeconomic stratification underscores that policy design must transcend the homogenous label of "women" to address the distinct needs of subgroups like young low-income mothers and highly educated professionals.

**2.2 Spatio-Temporal Context Heterogeneity**

The dynamic nature of spatiotemporal contexts renders women's nighttime ride-hailing usage intention highly situation-dependent, manifesting across three core dimensions: temporal rhythms, spatial characteristics, and service scenarios. Temporally, the period 22:00-02:00, characterized by a 40% peak in crime rates (National Crime Records Bureau, 2021), constitutes a high sensitivity window, triggering a 35% surge in order cancellations. In contrast, evening hours (18:00-22:00), driven by rigid social/leisure demands, reflect greater sensitivity to emotional benefits over safety concerns (Wang et al., 2023). Spatial heterogeneity in women's ride-haling behavior arises from the interaction between regional safety and micro-environmental features. At the macro level, capital cities in India exhibit a 5.45% higher ride-hailing frequency, attributed to improved infrastructure such as street lighting (Gupta et al., 2024). Conversely, high crime rates in non-capital northern cities (e.g., Varanasi) directly suppress usage. At the micro level, pickup/drop-off locations are critical—crime rates in secluded alleys are 70% higher than on main roads, and low-visibility areas such as underpasses trigger strong avoidance behavior, substantially reducing demand (Peng et al., 2022). Service scenarios shape perceived risk through co-traveler composition and vehicle configuration. Women's acceptance of ride-pooling varies significantly by the gender of co-riders, with safety ratings plummeting to 2.01/5 when sharing with unknown men, compared to 4.59/5 with female friends (Meshram et al., 2020). The confined vehicle space amplifies sensitivity to driver behavior, with safety perceptions declining by 60% following detours or inappropriate conversation (Acheampong, 2021). Notably, technology adaptability buffers spatiotemporal risks: real-time GPS tracking increases usage intention by 25% in high-crime zones, and smart streetlight-integrated route planning avoids 67% of high-risk segments (Omar et al., 2022; Borres et al., 2023). These spatiotemporal heterogeneities highlight the need to improve ride-hailing safety within a "travel chain" framework, requiring end-to-end optimization from pedestrian access points to destination environments to effectively address women's nighttime mobility challenges.

**2.3 Psychological-Behavioral Heterogeneity**

Women's psychological cognition and behavioral strategies are fundamental drivers of nighttime ride-hailing usage intention, with marked heterogeneity across risk perception frameworks, protective behavior patterns, and cultural imprints. Risk perception is shaped by prior experiences and cultural context: sexual harassment victims exhibit tripled vigilance towards driver characteristics and amplify avoidance through *hypervigilance* mechanisms (Greenwald, 2012). Conversely, women in conservative regions (e.g., Rajasthan), where patriarchal norms frame nighttime mobility as socially deviant, exhibit usage rates that are only one-third those observed in metropolitan areas (Gargiulo et al., 2020). Protective behaviors among women manifest as either instrumental or ritualized strategies. Instrumental strategies leverage technology interventions; 87% of users employ trip-sharing features (Gardner et al., 2017), and mobile navigation apps are rated 4.02/5 in enhancing perceived safety (Chen & Lu, 2021). Ritualized strategies act as psychological defenses, including fake phone calls (53%), adjusting attire (50%), and feigning confidence (68%). While these may alleviate anxiety, they often lead to behavioral compromises (Whitzman, 2007; Chowdhury & Van Wee, 2020). Cultural imprints further differentiate group responses: Western women emphasize the right to independent exploration and are willing to pay higher safety premiums (€5-20/trip), whereas Asian women, influenced by collectivist traditions, tend to terminate trips early and report crimes at significantly lower rates (12%) (Mellgren et al., 2018; Yang et al., 2018). Crucially, platform trust acts as a non-linear moderator of women's nighttime ride-hailing behavior. Infrequent users (<1 trip/month) prioritize traffic safety, while frequent users (>10 trips/month) emphasize data privacy. A notable "safety-indifferent" segment (47%) remains relatively risk-insensitive yet exhibits strong price sensitivity and vulnerability to negative event shocks (IFC, 2018; Tang et al., 2021). The psycho-behavioral heterogeneity underscores that women's nighttime travel choices reflect a negotiation between safety needs and agency demands, necessitating policy interventions that integrate both internal cognition factors and external environment conditions.

## 3. DATA AND SAMPLE DESCRIPTION

From April 11 to May 11, 2025, a questionnaire survey was conducted via the Wenjuanxing platform (https://www.wjx.cn/), targeting female passengers with nighttime ride-hailing experience through random sampling. To prevent duplicate entries, the platform tracked IP addresses and usernames. However, no personally identifiable information was collected or accessible. Participants were informed that the survey was anonymous, excluded any private data, and was exclusively for academic research. To encourage response rates, a modest incentive of 1 to 5 RMB was provided upon survey completion.

The survey instrument comprised four sections: (a) a revealed preference (RP) module capturing respondents' nighttime ride-hailing travel experiences; (b) a situational perception module employing five-point Likert-scale items to assess psychological attitudes toward nighttime ride-hailing; (c) a stated preference (SP) component evaluating preferences for key ride-hailing attributes (e.g., acceptable waiting time, driver rating requirements); (d) basic sociodemographic information (e.g., age, occupation, and residential location).

### 3.1 Observable Variables Data Statistics

To ensure robust regional representation, respondents were recruited from 29 provinces across

China. The regional distribution of the sample is illustrated in **Fig. 1**. Respondents from Guangdong (11.23%), Jiangsu (10.31%), and Shanghai (9.39%) constituted relatively higher proportions, followed by Shandong (7.37%) and Henan (5.34%). Overall, the spatial distribution represents a reasonably balanced and representative national sample.

A total of 543 female participants submitted valid questionnaires. **Table 1** presents the distribution of key variables. Categorical variables such as age, education level, and city of residence generally followed normal distributions, indicating reasonable representativeness across demographic categories. In terms of occupation, students (24.31%) and employees of enterprises/corporations (44.57%) constituted the majority. Given that students typically do not report personal income, self-reported consumption level was used as a proxy for income. Additional sociodemographic indicators include possession of a driver's license (58.56%) and private car ownership (30.76%), presence of children in the household (49.17%), and cohabitation with elderly individuals (56.17%). Regarding nighttime ride-hailing usage, respondents reported relatively balanced trip purposes across four predefined categories (night work, go home, entertainment, and shopping). The most common monthly ride-hailing frequency was 1-5 trips (32.78%), with the highest proportion of monthly expenditure falling 50-100 RMB range (28.91%).

Approximately 34.44% of female respondents reported experiencing unpleasant incidents during ride-hailing trips. These incidents included verbal or physical harassment, traffic accidents, route disputes, and fare disagreements [see **Fig. 2(a)** for details]. Although the incidence of sexual harassment was comparatively low (7.49%), its presence underscores a persistent and critical safety concern, particularly salient in the context of nighttime travel. Service-related disputes (route disputes: 26.74%, fare disagreements: 21.93%) were more commonly reported, reflecting systemic deficiencies in trip transparency (e.g., detours) and fare reasonableness (e.g., surge pricing), which increase the potential for conflict between female passengers and drivers.

Furthermore, the results indicate a generally low willingness among women to use ride-pooling services at night, with 73.30% of respondents indicating reluctance [see **Fig. 2(b)**]. In addition, 69.24% perceived nighttime ride-hailing fares as comparatively high, suggesting price sensitivity remains a significant barrier. Notably, only 30.20% of respondents reported having encountered a female driver during nighttime trips, implying that the vast majority, nearly 70%, interacted exclusively with male drivers. While these findings reflect prevailing conditions within the ride-hailing industry, they also underscore persistent constraints that inhibit women's nighttime ride-hailing usage, particularly in relation to service modality, pricing, and gendered perceptions of safety. To construct a panel dataset suitable for choice modeling, two scenario variables were defined: Ride-hailing Waiting Time (WT) and Ride-hailing Travel Time (TT), each with five discrete levels. WT levels: <5 min, 6-10 min, 11-15 min, 16-20 min, >20 min. TT levels: <10 min, 11-20 min, 21-30 min, 31-40 min, >40 min. Based on each respondent's state thresholds for acceptable WT and TT, a personalized choice scenario was constructed. Specifically, if a respondent reported willing to use nighttime ride-hailing only if WT <5 min and TT <10 min, then the decision variable for *only* that specific scenario combination (WT<5 min & TT<10 min) was coded as 1 (representing "willing"), while the remaining nine scenario combinations were coded as 0 (representing "unwilling"). This approach created ten distinct choice scenarios per respondent, resulting in a total of 5,430 observations (543 respondents × 10 scenarios) across 543 respondents, which was subsequently utilized for the LC-ICLV model estimation.

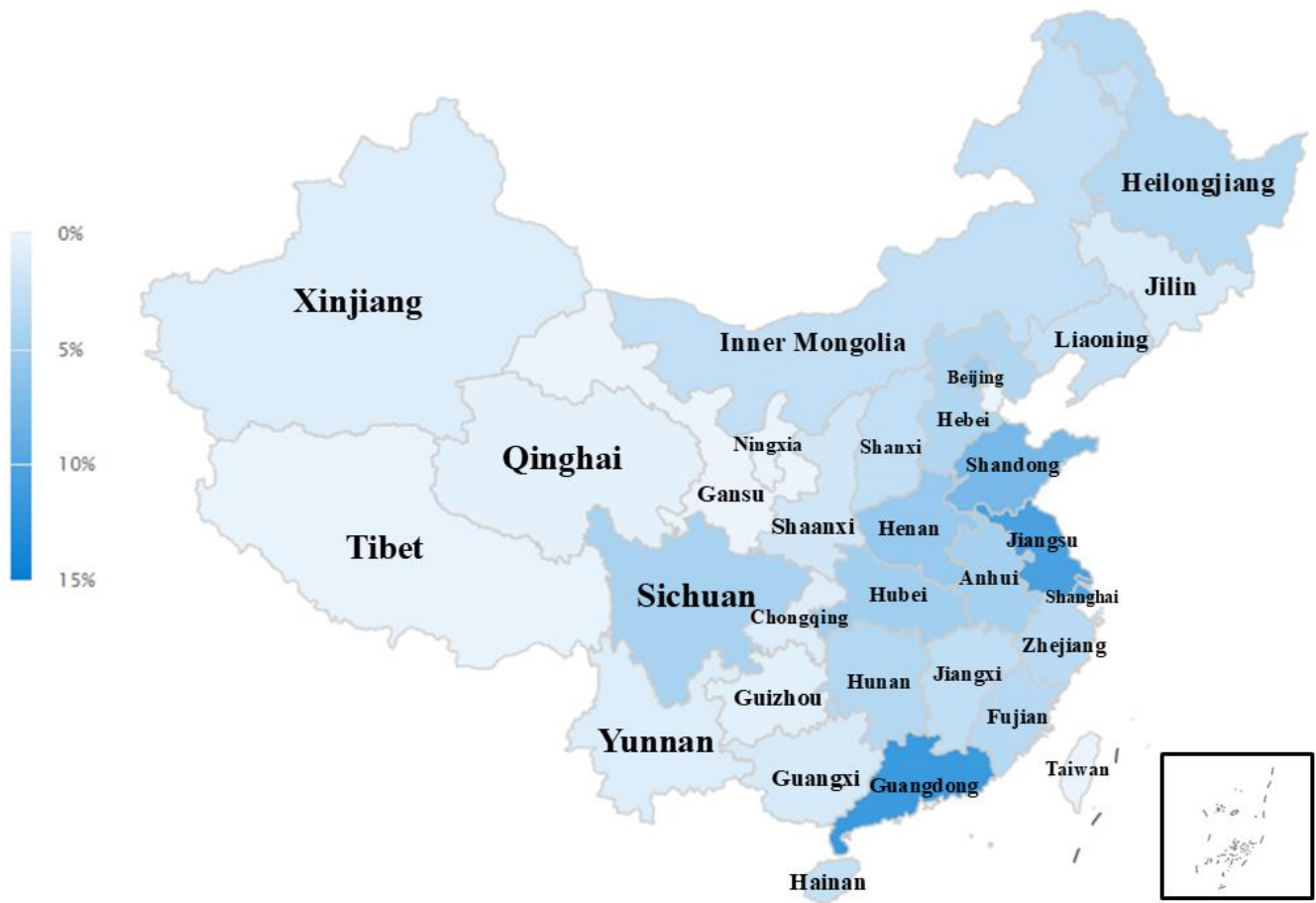

**Fig. 1.** Regional distribution of the sample.

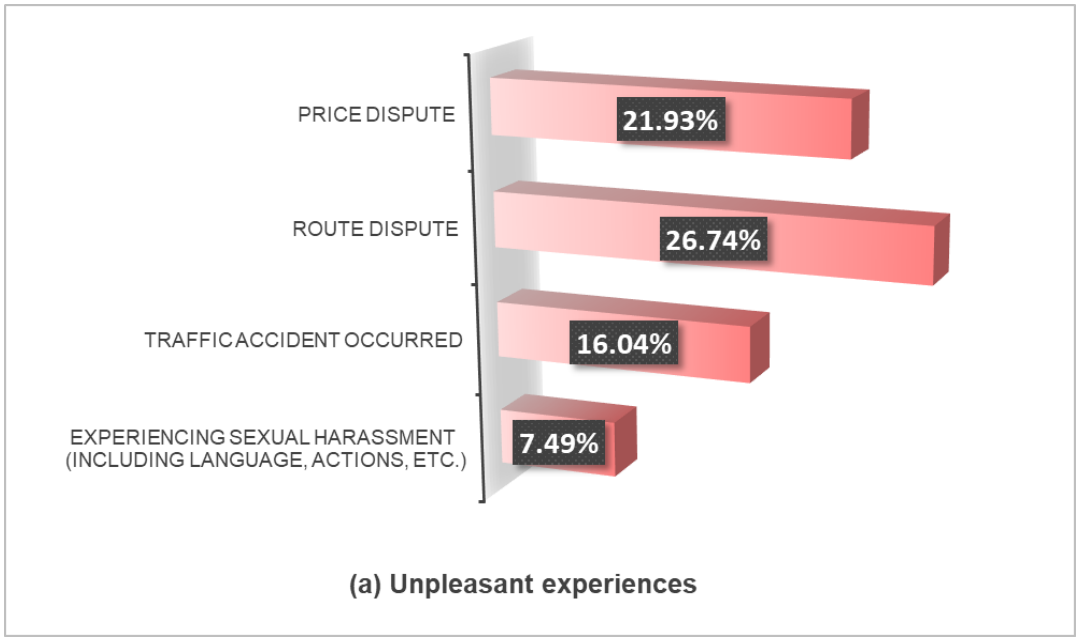

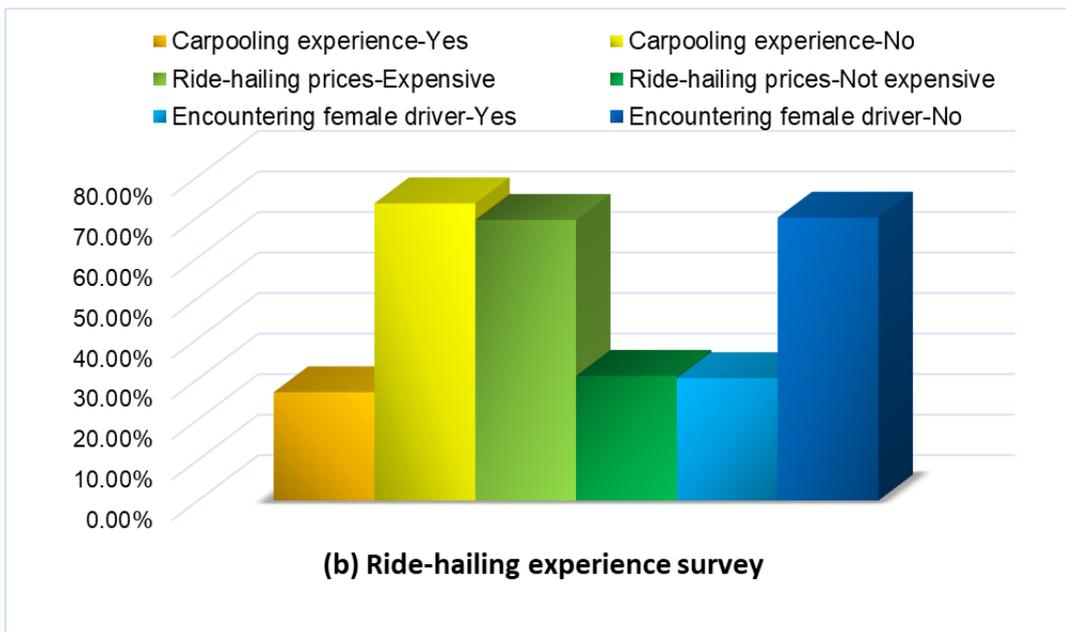

**Fig. 2.** Statistics of women's nighttime ride-hailing willingness and experiences

**Table 1.** Sample description

| Variables | Categories | Count | Percentage | Variables | Categories | Count | Percentage |
|---|---|---|---|---|---|---|---|
| Age | <25 years | 166 | 30.57% | Unpleasant experience | Yes | 187 | 34.44% |
| | 26-40 years | 220 | 40.52% | | No | 356 | 65.56% |
| | 41-50 years | 105 | 19.34% | Trip purpose | Night work | 181 | 33.33% |
| | >50 years | 52 | 9.58% | | Go home | 213 | 39.23% |
| Occupation | Student | 132 | 24.31% | | Entertainment | 206 | 37.94% |
| | Staff and workers of the enterprise | 242 | 44.57% | | Shopping | 223 | 41.07% |
| | Government or public institution personnel | 60 | 11.05% | Travel together | Yes | 228 | 41.99% |
| | Individual business owners | 109 | 20.07% | | No | 315 | 58.01% |
| Education | High school and below | 91 | 16.76% | Carpooling experience | Yes | 145 | 26.70% |
| | College | 306 | 56.35% | | No | 398 | 73.30% |
| | Postgraduate | 101 | 18.60% | Ride-hailing prices | Expensive | 376 | 69.24% |
| Income | <2000 RMB | 124 | 22.84% | | Not expensive | 167 | 30.76% |
| | 2000-5000 RMB | 116 | 21.36% | Ride-hailing on large platforms | Yes | 375 | 69.06% |
| | 5001-10000 RMB | 156 | 28.73% | | No | 168 | 30.94% |
| | 10001-30000 RMB | 75 | 13.81% | Encountering female driver | Yes | 164 | 30.20% |
| | >30000 RMB | 72 | 13.26% | | No | 379 | 69.80% |
| Residential city | First-tier cities | 120 | 22.10% | Frequency | Almost not used | 75 | 13.81% |
| | Second-tier cities | 144 | 26.52% | | 1-5 times per month | 178 | 32.78% |
| | Third-tier cities | 125 | 23.02% | | 6-10 times per month | 143 | 26.34% |
| | Fourth-tier cities | 81 | 14.92% | | 11-15 times per month | 83 | 15.29% |
| | Below fourth-tier cities | 73 | 13.44% | | >15 times per month | 64 | 11.79% |
| Driving License | Yes | 318 | 58.56% | Expenditure on ride-hailing per month | <50 RMB | 108 | 19.89% |
| | No | 225 | 41.44% | | 50-100 RMB | 157 | 28.91% |
| Private vehicle | Yes | 376 | 69.24% | | 101-300 RMB | 122 | 22.47% |
| | No | 167 | 30.76% | | >300 RMB | 90 | 16.57% |
| Marital Status | Yes | 313 | 57.64% | Maximum acceptable markup price | | | |
| | No | 230 | 42.36% | Maximum ride-hailing discount | | | |
| Children aged 0-12 | Yes | 267 | 49.17% | Driver score | Continuous variable | 543 | - |
| | No | 276 | 50.83% | Ride-hailing waiting time | | | |
| Elderly people aged 60 and above | Yes | 305 | 56.17% | Ride-hailing in-vehicle time | | | |
| | No | 238 | 43.83% | | | | |
| Relatives | Engaged in ride-hailing services | 236 | 43.46% | | | | |
| | Not engaged in ride-hailing services | 307 | 56.54% | | | | |

**3.2 Latent Variable Data Statistics**

To gain specific insights into respondents' experiences with ride-hailing services, this study employed a five-point Likert scale (ranging from 1 = "Strongly Disagree" to 5 = "Strongly Agree") to assess female passengers' perceptions across various scenarios. Three latent constructs were developed: Perceived Risk, Perceived Pleasure, and Perceived Safety, measured using specific items grouped into the following seven dimensions: Individual Perceived (IP1-IP6), Ride-hailing Environment (TE1-TE4), Interior Environment (IE1-IE4), Interior Behavior (IB1-IB4), Driver Image (DI1-DI4), Driver Behavior (DB1-DB6), and Platform Behavior (PB1-PB4). Descriptive statistics (see **Tables 2–4**) show that item means generally fell between 3 and 4, aligning with survey design expectations. Cronbach's alpha coefficients for all constructs exceeded 0.8, indicating high internal consistency and construct reliability (Urbach and Ahlemann, 2010).

**Table 2.** Analysis and testing of perceived risk scale

| Constructs | Mean | SD | α |
|---|---|---|---|
| Individual Perception (IP) | | | 0.898 |
| IP1: My personality is quite sensitive and suspicious | 3.29 | 1.10 | |
| IP2: I am a person who lacks a sense of security | 3.30 | 1.13 | |
| IP3: I feel scared and fearful when I go out at night | 3.33 | 1.16 | |
| IP4: The bad ride-hailing experiences of family and friends make me feel scared | 3.36 | 1.16 | |
| IP5: Once I am subjected to harassment or other unpleasant experiences, they will leave a vivid impression on me | 3.38 | 1.19 | |
| IP6: I have ongoing concerns about negative social cases involving ride-hailing services, such as assaults or murders | 3.32 | 1.17 | |
| Ride-hailing Environment (TE) | | | 0.857 |
| TE1: I will try my best to avoid taking ride-hailing in dimly lit areas | 3.36 | 1.13 | |
| TE2: I don't want to stay in areas with poor hygiene for a long time | 3.34 | 1.13 | |
| TE3: I tend to feel gloomy when I am in a green belt or a place with many trees | 3.35 | 1.12 | |
| TE4: In bad weather, I prefer to use ride-hailing services and leave as soon as possible | 3.38 | 1.13 | |

**Table 3.** Analysis and testing of perceived pleasure scale

| Constructs | Mean | SD | α |
|---|---|---|---|
| Interior Environment (IE) | | | 0.860 |
| IE1: Turning on the interior lights while riding at night gives me peace of mind | 3.40 | 1.11 | |
| IE2: The absence of any odor in the car will enhance my satisfaction with the ride | 3.51 | 1.13 | |
| IE3: Comfortable seats in the car will enhance my riding experience | 3.52 | 1.16 | |

| | Mean | SD | α |
|---|---|---|---|
| IE4: The cleanliness of the exterior and interior environment of car will reduce my anxiety about taking ride-hailing services at night | 3.53 | 1.14 | |
| Interior Behavior (IB) | | | 0.845 |
| IB1: Playing phone during the ride can alleviate my anxiety to some extent | 3.36 | 1.12 | |
| IB2: Wearing headphones and listening to music during the ride can distract my attention | 3.33 | 1.14 | |
| IB3: If my phone is fully charged in the car, it can give me a sense of security | 3.38 | 1.14 | |
| IB4: I will sit on the same side as the driver to ensure my privacy and safety | 3.39 | 1.15 | |

**Table 4.** Analysis and testing of perceived security scale

| Constructs | Mean | SD | α |
|---|---|---|---|
| Driver Image (DI) | | | 0.847 |
| DI1: The driver's strange accent makes me feel unfamiliar | 3.35 | 1.19 | |
| DI2: The smell of smoke on the driver's body makes me feel uncomfortable | 3.40 | 1.18 | |
| DI3: I feel scared when the driver is over fifty years old | 3.38 | 1.17 | |
| DI4: A driver who looks scary makes me feel unsafe | 3.34 | 1.14 | |
| Driver Behavior (DB) | | | 0.905 |
| DB1: Driving with one hand and operating the navigation system on my phone makes me feel very unsafe | 3.34 | 1.21 | |
| DB2: If the driver drives too fast, I will feel very anxious | 3.40 | 1.18 | |
| DB3: I will be very scared if the driver doesn't follow the navigation | 3.49 | 1.22 | |
| DB4: The driver often talks to me, which makes me very irritable | 3.31 | 1.17 | |
| DB5: The driver staring at me through the rearview mirror feels uncomfortable | 3.45 | 1.17 | |
| DB6: The driver making phone calls while driving makes me feel very unsafe | 3.49 | 1.15 | |
| Platform Behavior (PB) | | | 0.875 |
| PB1: The regular safety self-inspection and public promotion of ride-hailing platforms will promote my use of ride-hailing | 3.28 | 1.22 | |
| PB2: Drivers can feel at ease displaying their personal ID card at the front of the car | 3.37 | 1.20 | |
| PB3: More female drivers joining ride-hailing platforms will feel that travel is becoming increasingly safe | 3.41 | 1.23 | |
| PB4: The diverse protection measures provided by the platform for users (such as emergency contacts, one-click alarm) make me feel comprehensive | 3.33 | 1.17 | |

As shown in **Table 5**, the measurement model demonstrates strong psychometric performance. Convergent Validity is supported, with all Average Variance Extracted (AVE) values exceeding the 0.5 threshold. Composite Reliability (CR) is also confirmed, with all CR values exceeding the 0.8 benchmark. Together, these results indicate that all seven latent variable dimensions possess robust convergent validity and composite reliability. Additionally, Discriminant Validity is likewise established, with the square root of each construct's AVE value exceeding its correlations with other constructs, confirming adequate distinction among the latent dimensions.

**Table 5.** Results of convergence validity, combination reliability, and discriminant validity tests

|  | IP | TE | IE | IB | DI | DB | PB | CR |
|---|---|---|---|---|---|---|---|---|
| IP | **0.596** |  |  |  |  |  |  | 0.899 |
| TE | 0.461 | **0.600** |  |  |  |  |  | 0.857 |
| IE | 0.396 | 0.479 | **0.609** |  |  |  |  | 0.861 |
| IB | 0.367 | 0.454 | 0.462 | **0.580** |  |  |  | 0.846 |
| DI | 0.525 | 0.456 | 0.525 | 0.497 | **0.580** |  |  | 0.847 |
| DB | 0.469 | 0.532 | 0.494 | 0.454 | 0.501 | **0.615** |  | 0.905 |
| PB | 0.473 | 0.531 | 0.468 | 0.396 | 0.552 | 0.537 | **0.638** | 0.876 |
| $\sqrt{AVE}$ | 0.772 | 0.775 | 0.780 | 0.762 | 0.762 | 0.784 | 0.799 |  |

## 4．METHODS

### 4.1 Research Framework and Hypotheses

**Fig. 3** details the methodological framework of this study. This paper employs the Latent Class Integrated Choice and Latent Variable (LC-ICLV) model to elucidate the influencing factors and heterogeneous decision-making behaviors underlying women's willingness to use ride-hailing services at night by jointly considering the effects of individual socio-demographic characteristics, ride-hailing service attributes, and latent attitudinal/perceptual variables. The approach first utilizes a Latent Class Model (LCM) to segment the overall population into distinct latent subgroups exhibiting heterogeneous preference patterns. Subsequently, within each identified subgroup, the ICLV model is independently applied for simultaneous estimation: a structural model constructs the latent variables influencing individual choices, a measurement model validates these latent variables using indicator variables, and then the estimated latent variables are incorporated into the utility function of the choice model (with a mixed Logit kernel) as key explanatory factors. This methodological framework not only uncovers the underlying structure of preference heterogeneity but also precisely quantifies, at the group level, the driving effect of psychological factors like attitudes and perceptions on actual choice behavior.

Two critical methodological points warrant further clarification:

I. Choice Model Specification: Given the binary nature of the choice decision in this study (the dependent variable is whether a woman is willing or unwilling to use ride-hailing at night), a Binary Logit (BL) model is adopted as the choice model component within the ICLV framework.

Furthermore, to better explore the behavioral heterogeneity across explanatory variables among female subgroups with distinct preference structures, the BL model is extended by randomizing the coefficients of its explanatory variables. This extension transforms the BL model into a Mixed Logit (ML) specification within the ICLV structure for each latent class. By incorporating random parameters, the model accounts for unobserved preference variation within classes, thereby offering a more nuanced understanding of the complexity underlying women's ride-hailing decision-making.

II. Measurement Model Enhancement: As introduced earlier, and drawing on key insights from the Generalized Heterogeneous Data Model (GHDM, Bhat 2015), this study relaxes the default assumption of continuous indicator variables in the measurement model. Instead, the indicators are treated as ordered categorical variables. Accordingly, the measurement component of the ICLV model employs the Ordered Probit Model specification for probability estimation. This modification enhances the realism and flexibility of the model, allowing for a more appropriate representation of the underlying latent constructs.

The underlying research assumptions of the proposed LC-ICLV model encompass the following three tenets:

I. Utility Maximization: Female individuals seek to maximize their utility or satisfy their preferences when making choices. That is, their final decision (to use or not use ride-hailing at night) results from maximizing perceived benefits.

II. Significant Intra-Group Heterogeneity: The behavioral decisions of the female population exhibit substantial internal heterogeneity. To capture this complexity, the model integrates group-level segmentation via LCM and individual-level random effects via ML specification within each class. This dual-layer structure accounts for both *inter-group* and *intra-group* heterogeneity, thereby providing insights into why different women or women within the same latent subgroup might make divergent choices under similar circumstances.

III. Structural Causal Pathways: Explicit structural relationships are specified between the latent variables and the decision to use ride-hailing services at night. Specifically, these latent constructs exert a causal influence on the parameters of the choice model, thereby mediating the effects of underlying perceptions and attitudes on the observed choice outcomes.

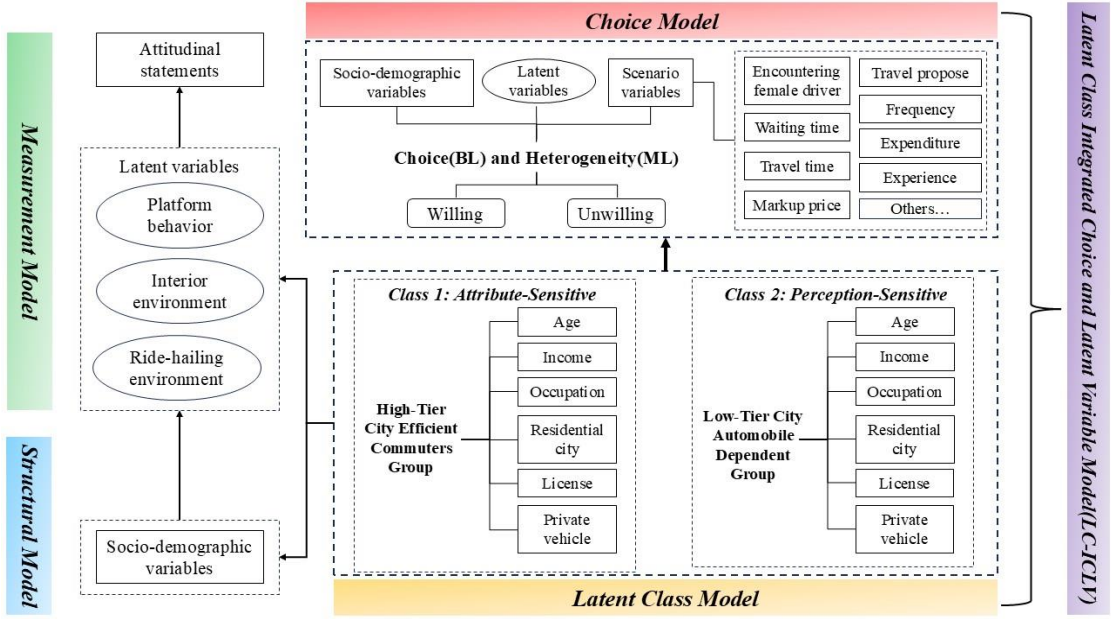

**Fig. 3.** Conceptual modeling framework of the LC-ICLV model.

**4.2 Modeling of LC-ICLV**

*4.2.1 Latent Class Model (LCM)*

The probability that individual *n* belongs to latent class *q* is defined using a multinomial logit model as follows:

$$\pi_{nq} = (C_n = q \mid Z_n; \gamma_q) = \frac{\exp(\gamma_q^T Z_n)}{\sum_{q'=1}^{Q} \exp(\gamma_{q'}^T Z_n)}, q = 1,2,\ldots,Q \tag{4.1}$$

where $Z_n$ denotes an $(M \times 1)$ vector of socio-economic characteristics for female individual $n$, and $\gamma_q$ is the corresponding $(M \times 1)$ parameter vector for class $q$. The first class is normalized as the reference category by setting $\gamma_1 = 0$.

*4.2.2 Latent Class Integrated Choice and Latent Variable Model(LC-ICLV)*

For each latent class *q*, the structural model of the ICLV framework is specified in matrix form as follows:

$$X_{nq}^* = \Lambda_q X_n + \omega_q, \omega_q \sim N(0, \Psi_q) \tag{4.2}$$

where $X_{nq}^*$ represents a $(G \times 1)$ vector of latent variables that influence women's nighttime choices within latent class $q$. The vector $X_n$ denotes the $(K \times 1)$ dimensional explanatory variables, including individual socio-economic characteristics and ride-hailing service attributes. $\Lambda_q$ signifies the $(G \times K)$ dimensional parameter matrix to be estimated, capturing the relationship between latent variables and observed covariates. $\omega_q$ stands for a $(G \times 1)$ dimensional random disturbance term, assumed to have a mean of zero and a covariance matrix $\Psi_q$, where $\Psi_q$ is of dimension $(G \times G)$.

Based on **Eq. (4.2)**, the probability density function (PDF) of the structural model is derived as follows:

$$f_{X^*,q}(X^*_{nq} \mid X_n) = (2\pi)^{-\frac{G}{2}}|\Psi|^{-\frac{1}{2}}\exp\left[-\frac{1}{2}(X^*_{nq} - \Lambda_q X_n)^T \Psi_q^{-1}(X^*_{nq} - \Lambda_q X_n)\right] \quad (4.3)$$

Next, the measurement component of the ICLV model is defined in a matrix expression. The indicator variables are treated as ordinal categorical variables. Accordingly, an Ordered Probit model is employed to derive the matrix expression for the continuous latent variable vector $I^*_{nq}$, which underlies the observed ordinal indicator vector $I_{nq}$ for individual $n$ in latent class $q$:

$$I^*_{nq} = D_q X^*_{nq} + \xi_q + v_q, v_q \sim N(0, \Theta_q) \quad (4.4)$$

where $I^*_{nq}$ represents the $(H \times 1)$ dimensional unobservable continuous latent variable vector for individual $n$ in latent class $q$. $D_q$ is a $(H \times G)$ dimensional matrix of parameters to be estimated. The term $\xi_q$ is an $(H \times 1)$ vector of intercepts and $v_q$ is an $(H \times 1)$ random disturbance vector, assumed to have a mean of zero and a variance-covariance structure defined by $\Theta_q$, an $(H \times H)$ diagonal matrix.

Therefore, the conditional probability of observing the ordered indicator vector $I_n = m = (m_1, \dots, m_H)^T$ is given by:

$$P(I_{nq} = m \mid X^*_{nq}) = \prod_{h=1}^{H} \Phi\left(\frac{\tau_{h,m_h,q} - D_{hq}X^*_{nq}}{\sqrt{\Theta_{hq}}}\right) - \Phi\left(\frac{\tau_{h,m_h-1,q} - D_{hq}X^*_{nq}}{\sqrt{\Theta_{hq}}}\right) \quad (4.5)$$

where $D_{hq}$ is the *h-th* row of the parameter matrix $D_q$ to be estimated. $\tau_{h,m_h,q}$ represents the *m-th* threshold parameter of *h-th* indicator in latent class $q$ $(\tau_{h,0,q} = -\infty, \tau_{h,M,q} = +\infty)$; $\Phi(\cdot)$ is the cumulative distribution function of the standard normal distribution.

Accordingly, the PDF of the measurement equation can be expressed as follows:

$$f_{I,q}(I_n \mid X^*_{nq}) = \prod_{h=1}^{H} P(I_{nh} = m_h \mid X^*_{nq}) \quad (4.6)$$

According to the random utility maximization (RUM) theory, the final choice of individual $n$ can be expressed as:

$$y_{jn} = \begin{cases} 1, & \text{When } U_{nj} = \max_{j' \in J}\{U_{j'n}\} \\ 0, & otherwise \end{cases} \quad (4.7)$$

where $y_{jn}$ represents the choice outcome of traveler $n$ selecting alternative $j$, and $J$ is the total number of available alternatives.

Next, the utility function in the choice model of the ICLV framework is presented as follows:

$$U_{nq} = E_q X_n + \Gamma_q X^*_{nq} + \varepsilon_{nq} \quad (4.8)$$

where $U_{nq}$ is the $(J \times 1)$ dimensional utility vector of traveler $n$ within latent class $q$; $E_q$ and $\Gamma_q$ are $(J \times K)$ and $(J \times G)$ dimensional parameter matrices, respectively; $\varepsilon_{nq}$ represents the random error component, assumed to follow an independent and identically distributed (i.i.d.) Type I extreme value distribution.

The probability of traveler *n* choosing alternative *j* in the ICLV model is derived as follows:

$$P_{nj|q} = \frac{\exp(E_q X_n + \Gamma_q X_{nq}^*)}{\sum_{i \in J} \exp(E_q X_n + \Gamma_q X_{nq}^*)} \quad (4.9)$$

It should be noted that when **Eq. (4.7)** is incorporated into the BL model, the parameter vectors $E_q$ and $\Gamma_q$ are both fixed parameters. However, with the introduction of Mixed Logit (ML) models, these parameters are specified as random variables following a certain distribution rather than fixed values. In this study, the random parameters in the ML model are assumed to follow the commonly adopted normal distribution.

Furthermore, the probability density function of the choice outcome is expressed as follows:

$$f_{y,q}(y_n \mid X_{nq}^*) = \prod_{j \in J} (P_{nj \mid q})^{y_{jn}} \quad (4.10)$$

Finally, based on **Eq. (4.1)**, **(4.3)**, **(4.6)**, and **(4.10)**, the joint probability density function of the observed data $(y_n, I_n)$ is obtained by integrating over the latent variable $X_{nq}^*$:

$$f_{y,I}(y_n, I_n \mid X_n; \theta) = \sum_{q=1}^{Q} \int f_{X^*,q}(X_{nq}^* \mid X_n) \cdot f_{I,q}(I_n \mid X_{nq}^*) \cdot f_{y,q}(y_n \mid X_{nq}^*) \, dX_{nq}^* \quad (4.11)$$

where the set of parameters to be estimated is $\theta = \left\{ \{\gamma_q\}_{q=2}^{Q}, \Lambda_q, \Psi_q, D_q, \xi_q, \Theta_q, \tau_q, E_q, \Gamma_q \right\}_{q=1}^{Q}$.

*4.2.3 Parameter Estimation of LC-ICLV Model*

Parameter estimation of the LC-ICLV model is conducted using the Monte Carlo simulation method. Given the high dimensionality of latent variables in this study, Halton sequences were employed as the Monte Carlo sampling technique due to their superior efficiency in handling high-dimensional integration. To ensure the stability and reliability of the parameter estimates, each estimation was based on approximately 2,000 simulation draws.

Firstly, for each latent class *q*, *R* independent and identically distributed samples of the latent variables $X_{nq}^{*(1)}, X_{nq}^{*(2)}, \ldots, X_{nq}^{*(R)}$ are drawn from the probability density function of the structural model, which can be expressed mathematically as:

$$\left\{ X_{nq}^{*(r)} \right\}_{r=1}^{R} \sim N(\Lambda_q X_n, \omega_q) \quad (4.12)$$

Subsequently, based on the results from **Eq. (4.11)**, the Monte Carlo approximation of the joint probability density within each latent class is computed as:

$$f_{y,I}(y_n, I_n \mid X_n; \theta) \approx \sum_{q=1}^{Q} \pi_{nq} \left[ \frac{1}{R} \sum_{r=1}^{R} f_{y,q}(y_n \mid X_{nq}^{*(r)}) \cdot f_{I,q}(I_n \mid X_{nq}^{*(r)}) \right] \quad (4.13)$$

Finally, the logarithmic expression of the maximum likelihood function for parameter estimation in the LC-ICLV model is expressed as follows:

$$\max_{\theta} \sum_{n=1}^{N} \ln \left\{ \sum_{q=1}^{Q} \pi_{nq} \left[ \frac{1}{R} \sum_{r=1}^{R} f_{y,q}(y_n \mid X_{nq}^{*(r)}) \cdot f_{I,q}(I_n \mid X_{nq}^{*(r)}) \right] \right\} \quad (4.14)$$

# 5. MODEL RESULTS AND ANALYSIS

## 5.1 Selection of the Number of Classes

Selecting an appropriate number of latent classes is crucial, as it substantially influences the interpretability of model results. However, determining the criteria for the optimal number of classes remains a challenging and evolving domain of research (Weller et al., 2020). Nonetheless, in practice, researchers typically rely on statistical criteria, among which the Bayesian Information Criterion (BIC) (Nylund et al., 2007) is the most widely applied metric. Additional criteria such as the Akaike Information Criterion (AIC), Finite Sample Corrected AIC (CAIC), and Hannan-Quinn Information Criterion (HQIC) are also used. Lower values of these criteria indicate better model fit and guide the selection of the preferred model specification.

However, the objective of this study extends beyond identifying the model with the best predictive performance; it also seeks to uncover deeper insights into the heterogeneity of women's preferences regarding nighttime ride-hailing service usage. To this end, we first examined the statistical criteria and subsequently evaluated whether the resulting classes exhibited clear behavioral interpretability. The analysis (as shown in **Table 6**) revealed that the two-class specification yielded the lowest values across multiple selection criteria, indicating superior statistical performance. In contrast, increasing the number of classes (Class ≥ 3) led to greater model complexity without a commensurate improvement in interpretability or fit.

Furthermore, this study considered the sample size across latent classes. A class comprising too few observations may undermine the model's goodness-of-fit and raise concerns regarding its statistical validity. Drawing on established criteria from previous scholars (Marsh et al., 2009; Ferguson et al., 2020), we excluded any model configurations in which a latent class contained less than 10% of the total sample size. As a result, models with more than 6 classes were ruled out. While a larger number of classes could theoretically capture more nuanced heterogeneity, preserving model parsimony is crucial to ensuring interpretability and practical utility (Kamakura & Russell, 1989). Notably, in the selected two-class model, the respondent distribution of respondents across classes was found to be balanced and reasonable. Considering all factors—statistical fit, class distinctiveness, adequate sample size per class, and interpretability—the two-class model specification was deemed to be the optimal solution.

**Table 6.** Quantitative fit of 1–6 latent class membership models.

| Class | LL | BIC | AIC | FIC | HQIC | Number of parameters | Class size |
|---|---|---|---|---|---|---|---|
| 1 | -3962.7 | 7809.0 | 7836.4 | 7836.5 | 7861.8 | 11 | 1 |
| **2** | **-3757.1** | **7643.8** | **7542.0** | **7542.2** | **7595.0** | **23** | **0.22/0.78** |
| 3 | -3757.2 | 7784.5 | 7553.5 | 7554.0 | 7634.1 | 35 | 0.27/0.32/0.41 |
| 4 | -3757.2 | 7878.2 | 7568.0 | 7568.8 | 7676.3 | 47 | 0.24/0.25/0.25/0.26 |
| 5 | -3757.3 | 7966.4 | 7577.0 | 7578.3 | 7712.9 | 59 | 0.32/0.16/0.14/0.21/0.17 |
| 6 | -3757.2 | 8069.3 | 7600.7 | 7602.6 | 7764.2 | 71 | 0.27/0.20/0.09/0.14/0.08/0.22 |

Note: LL = log-likelihood; AIC = Akaike Information Criterion; BIC = Bayesian Information Criterion; FIC= Finite Sample Corrected AIC; HQIC= Hannan-Quinn Information Criterion.

### 5.2 Estimation Results of LCM

The latent class analysis identified two distinct segments within the female respondent population, with Class 1 serving as the reference group. As shown in **Table 7**, individuals in Class 2 are primarily aged between 26-50 years ($\beta$=0.489, 0.392) and are significantly less likely to be students ($\beta$=-0.622), suggesting that this class corresponds to the core demographic of working-age, household commuters. In addition, members of Class 2 exhibit notable socio-spatial differences: they are more likely to reside in third-tier, fourth-tier, or lower-tier cities ($\beta$=0.432, 0.776), and are drawn from both middle-to-low and high-income brackets. These characteristics highlight the presence of substantial heterogeneity in nighttime ride-hailing preferences across socio-demographic and spatial dimensions, underscoring the importance of accounting for both urban hierarchy and lifecycle stage in modeling women's transportation choices.

A particularly noteworthy finding pertains to the opposing coefficients for possessing a driver's license ($\beta$=-0.136) and owning a private car ($\beta$=0.246). While seemingly contradictory, this pattern reflects profound socio-structural realities and underscores the complexity of interpreting vehicle access in lower-tier urban contexts. Three key mechanisms may account for this discrepancy.

I. Household sharing: Within multi-generational households, especially in third-tier, fourth-tier, and lower-tier cities, private cars are often registered under 1-2 licensed family members as shared household assets, yet routinely utilized by unlicensed members (e.g., spouses or elderly relatives).

II. Informal Vehicle Classification: Pervasive use of license-exempt quasi-motor vehicles (e.g., electric tricycles, senior mobility scooters) in these regions leads respondents to categorize them as "private cars" despite regulatory differences.

III. Economic Barriers to Licensing: Driver's license training costs in China (3000-5000 RMB), equivalent to 1-2 months' income in lower-tier cities, prompt some households to engage in "unlicensed driving" rather than invest in certification for all members. This paradox starkly illustrates how transportation behaviors mirror regional developmental disparities.

In contrast, the coefficient signs for Class 1 suggest a predominance of residents from first- and second-tier cities, characterized by a high proportion of middle-to-high-income earners ($\beta=-0.932$) and students ($\beta=0.622$). This profile suggests greater female labor market participation within this group. Moreover, the combination of higher driver's license possession but lower private car ownership implies increased reliance on well-developed public transit infrastructure (e.g., subways, buses) common in major urban centers, and may also potentially reflect stronger environmental consciousness among these women.

Based on the results of the category model and the above analysis, we define Class 1 as "High Tier City Efficient Commuters Group" and Class 2 as "Low Tier City Automobile Dependent Group". In summary, the distinct socioeconomic and geographic profiles delineated by this two-class segmentation provide a robust foundation for investigating the heterogeneity in women's nighttime ride-hailing usage intentions across these divergent subgroups.

**Table 7**. Estimation results of two-class latent class model.

|  | Class 1 High-Tier City Efficient Commuters Group | | Class 2 Low-Tier City Automobile-Dependent Group | |
| --- | --- | --- | --- | --- |
| Variables | Value | t-test | Value | t-test |
| Intercept | | | **0.397*** | 2.78 |
| Age(26-40 years) | | | **0.489*** | 4.47 |
| Age (41-50 years) | | | **0.392*** | 3.14 |
| Residential city (Third-tier cities) | | | **0.432*** | 3.81 |
| Residential city (Below fourth-tier cities) | Reference group | | **0.776*** | 4.56 |
| Income (5001-10000 RMB) | | | **-0.478*** | -3.84 |
| Income (10001-30000 RMB) | | | **-0.932*** | -6.6 |
| Occupation(Student) | | | **-0.622*** | -4.66 |
| License | | | **-0.136*** | -1.67 |
| Private car | | | **0.246*** | 2.27 |

Note: ***$p < 0.01$; **$p < 0.05$; *$p < 0.1$.

## 5.3 Estimation Results of the LC-ICLV Model

Three latent variables—Taxi Environment, Interior Environment, and Platform Behavior—were selected to represent the dimensions of Perceived Risk, Perceived Pleasure, and Perceived Safety, respectively. These constructs were incorporated into the structural, measurement, and choice model of the LC-ICLV framework for comprehensive analysis.

### 5.3.1 Structural Model

The estimation results of the structural model are shown in **Table 8**, and the following are the analysis contents of the two classification groups:

High-Tier City Efficient Commuters Group (Class 1): Although some variables were not statistically significant, most contributed to explaining the latent variable formation mechanisms for this group. Occupation and city of residence variables partially reflected women's perceived risk and perceived pleasure attitudes towards nighttime ride-hailing within the Taxi Environment and Interior Environment dimensions. Divergent coefficients for enterprise staff and workers ($\beta = -0.028$) versus individual business owners ($\beta=0.044$) suggest heterogeneous risk perceptions across occupations. Employees, potentially facing frequent overtime, may rely heavily on ride-hailing for late-night commutes, thereby normalizing the experience and reducing perceived risk. Significant positive coefficients for employees of enterprises and institutions ($\beta = 0.066$) and government personnel ($\beta = 0.147$) indicate that stable occupational groups place greater emphasis on riding experience. Residents of third-tier ($\beta = 0.054$) and fourth-tier cities ($\beta = 0.061$) reported higher perceived pleasure compared to the reference group (second-tier cities). Perceived safety regarding platform behavior was jointly shaped by income and occupation. Government personnel showed a significant negative association ($\beta = -0.072$), potentially reflecting a cautious stance toward platform compliance within this state-affiliated group, while other occupations show no significant effects. Compared to the reference income group (5000-10000 RMB), the low-income group (<2000 RMB, $\beta = 0.023$) exhibited greater trust in platform mechanisms, whereas the high-income group (>30000 RMB, $\beta = -0.060$) showed a significant negative association. This finding aligns with the Class 1 profile, where students, predominant within the high-income bracket *in this context*, tend to rely on and trust standardized platform safety features (e.g., trip sharing, emergency contacts) to mitigate travel uncertainties. Conversely, high-income respondents reliant on public transit may have limited ride-hailing exposure, reducing their engagement with platform safety measures. These findings are consistent with prior analysis and further support the validation of the Class 1 segmentation.

Low-Tier City Automobile-Dependent Group (Class 2): All three latent variables (perceived risk, pleasure, and safety) were significantly shaped by age, income, occupation, and city of residence. Both the youngest (<25 years, $\beta = 0.356$) and oldest (>50 years, $\beta = 0.482$) users exhibited significantly higher perceived risk associated with the traditional taxi environment, alongside stronger reliance on platform safety measures (β = 0.416 and β = 0.442, respectively). These patterns highlight heightened safety sensitivities at the age extremes, reflecting their unique mobility vulnerabilities during nighttime travel. Perceived pleasure within the Interior Environment was primarily driven by the middle-income group (2000-5000 RMB, β = 0.476), indicating greater expectations for improved in-vehicle comfort, cleanliness, and service quality in this economic segment. A critical finding concerns city tier effects: compared to second-tier cities, residents in first-tier (β = 0.203), third-tier (β = 0.199), fourth-tier (β = 0.245, and lower-tier cities (β = 0.246) all reported significantly higher perceived safety associated with platform safety measures. This signifies that even in more developed cities within Class 2, digital safety mechanisms (e.g., driver background checks, GPS tracking, emergency contacts) are particularly influential. Policy support for enforcing and standardizing these platform safety features—especially in less regulated markets—could improve women's confidence in ride-hailing services and promote equitable access to safe nighttime mobility.

*5.3.2 Measurement Model*

The measurement model quantifies the relationship between latent variables and their associated attitudinal indicators. As shown in **Table 9**, factor loadings capture the strength of association between each observed item and its underlying latent variable, while intercepts indicate the expected item response when the latent variable is 0. The standard deviations reflect the degree of response variability across respondents for each measurement item. All indicators for Class 1 and Class 2 were statistically significant, with positive factor loadings. This indicates that higher values of the latent constructs are associated with stronger agreement with the corresponding attitudinal statements, thereby confirming the internal consistency and validity of the measurement model across latent dimensions.

*5.3.3 Choice Model*

This study examined the influence of both ride-hailing attributes and latent attitudinal variables on women's nighttime ride-hailing willingness, employing the LC-ICLV model to investigate class-specific heterogeneity in decision-making behavior. Class-specific models were refined by iteratively removing statistically insignificant observed variables. However, considering the critical importance of latent variables in this study, all latent variables were retained irrespective of significance. The final estimation results of the choice model are presented in **Table 10**.

First, the intercept terms in the choice model offer insights into baseline preferences for nighttime ride-hailing across the two latent classes. For Class 1, the intercept is near zero and statistically insignificant (β = -0.017), indicating a minimal or ambiguous inherent tendency toward or against ride-hailing. This suggests that: (1) any intrinsic aversion is weak or negligible; (2) the included explanatory variables, both observed service attributes (e.g., discounts, waiting time) and latent variables (e.g., perceived safety), adequately capture the decision-making dynamics for Class 1; and (3) Class 1 individuals likely form their preferences through a relatively rational, context-dependent evaluation of ride-hailing characteristics. As such, they may be classified as "attribute-sensitive" users. In contrast, the intercept for Class 2 is significantly negative (β = -0.372), suggesting an underlying, systematic reluctance to engage in nighttime ride-hailing. This indicates that: (1) unobserved psychological or contextual factors not captured in the model may exert a strong negative influence on using ride-hailing services at night, resulting in a baseline inclination *against* usage; and (2) this class exhibits a deeper-seated, intrinsic aversion likely tied to fundamental safety concerns. Accordingly, Class 2 can be characterized as "perception-sensitive", with decisions more strongly shaped by internalized fears and attitudinal barriers. Together, these intercept effects underscore the heterogeneous nature of women's nighttime ride-hailing behavior across socio-demographic segments.

Furthermore, both Waiting Time (WT) and Travel Time (TT) exhibit significant negative coefficients in both classes, albeit driven by distinct mechanisms. For the "attribute-sensitive" Class 1, increases in WT (β = -1.700) and TT (β = -1.320) signify greater temporal and perceived monetary costs, naturally reducing the overall utility of nighttime ride-hailing. In contrast, for the "perception-sensitive" Class 2, prolonged WT (β = -1.230) and extended TT (β = -1.120) amplify discomfort associated with being alone in a vehicle with a stranger driver during nighttime hours. These time-related factors significantly heighten perceived vulnerability and reinforce internal safety concerns, thereby reducing subsequent ride-hailing usage intentions at night.

Analysis of latent variable effects reveals consistent sign patterns within each class, indicating internal consistency. For Class 1 ("attribute-sensitive"), the significant positive coefficient for Platform Behavior (β = 0.938) highlights the importance of safety assurances provided by ride-hailing platforms and regulators (e.g., strict driver background checks, real-time trip sharing/monitoring, one-touch emergency buttons) in promoting nighttime ride-hailing usage. This suggests a key policy direction for platforms and transportation authorities aiming to increase uptake among rational, service-oriented users. In Class 2 ("perception-sensitive"), the significant negative coefficient for Perceived Risk in Taxi Environment (β = -1.110) indicates that poor pickup/drop-off environment (e.g., dimly lit, isolated areas) substantially heightens fear and deters nighttime ride-hailing usage. Conversely, the positive coefficient for Interior Environment (β = 0.410) suggests that enhanced in-vehicle comfort (e.g., interior lighting, no unpleasant odors, comfortable seating) can effectively alleviate anxiety during the trip for "perception-sensitive" women, thereby increasing the utility of nighttime ride-hailing. Therefore, for this group, avoiding risky pickup environments and enhancing vehicle cleanliness are effective strategies to encourage adoption.

Examining other variables influencing nighttime ride-hailing decisions in Class 1: Consistent with Section 5.2, this group comprises residents of First and second-tier cities, middle-to-high income earners (5001-30000 RMB), students, and those with low private car ownership. Several factors significantly increase their willingness to use ride-hailing services at night: Night Shift (β = 0.506), Previous Ride-Pooling Experience (β = 0.294), and Platform Discounts (β = 0.112). This reflects their structural dependence on ride-hailing as a substitute for limited late-night public transit options. For night-shift workers in particular, ride-hailing often serves as the only feasible option. Additionally, as a price-sensitive group, especially students, fare reductions/discounts significantly enhance nighttime usage willingness. A noteworthy and somewhat counterintuitive result is the negative impact of Encountering a Female Driver (β = -0.320) on nighttime ride-hailing willingness. However, it can be explained: while a female driver might slightly reduce fear, it does not eliminate safety concerns. Existing stereotypes (such as the perception that male drivers are more experienced or safer) may reinforce hesitation, especially among "attribute-sensitive" users who evaluate service features pragmatically. Consequently, encountering a female driver does not necessarily increase willingness. This finding aligns with our previous findings (Wang et al., 2025) and refines the stereotype's origin: "attribute-sensitive" users make negative evaluations based on perceived risk associated with specific service features (like driver's gender), impacting their final decision.

Considering other variables influencing Class 2 decisions: As per Section 5.2, this "perception-sensitive" group is characterized by ages 26-50, residing in third- and fourth-tier cities or below, income below 5000 RMB or above 30000 RMB, and higher private car ownership. Several factors positively influence their nighttime ride-hailing usage: Possessing a Driver's License (β = 0.268), Trip Purpose: Entertainment (β = 0.284), and Trip Purpose: Going Home (β = 0.237). These findings point to nuanced behavioral dynamics: (1) driving experience likely enhances their sense of situational control—improving route awareness and the ability to assess driver behavior—thereby partially reducing fear related to personal safety; (2) the 26-50 age group often engages in nighttime entertainment and social activities that may involve alcohol consumption. Faced with the risk of drunk driving versus perceived risks of ride-hailing, the latter becomes the safer and more acceptable alternative; (3) strong family responsibilities (e.g., the need to return home for childcare) further reinforce nighttime ride-hailing, especially in urban settings where late-night public transit is limited or unavailable. Crucially, while these situations lead to usage, it doesn't negate their "perception-sensitive" nature. Rather, they reflect constrained choices shaped by necessity and risk trade-offs. Even when ride-hailing is used, safety concerns remain salient. For Class 2, ride-hailing is less a preferred option than a conditional or context-specific coping mechanism—employed only when perceived risks can be managed or when alternatives are absent.

The opposing coefficients for High Monthly Ride Frequency (>15) (β = -0.391) and High Monthly Ride Expenditure (>300 RMB) (β = 0.503) in Class 2 underscore significant intra-group heterogeneity in nighttime ride-hailing behavior. Frequent users may experience cumulative anxiety from repeated minor negative incidents (e.g., detours, uncomfortable interactions with drivers, or abrupt vehicle maneuvers), which—when compounded by nighttime safety concerns—suppress their willingness to use the service. Conversely, high monthly expenditure reflects a willingness to pay for "perceived safety." High-income users in Class 2 often choose premium services (e.g., premium cars), which offer better vehicles, stricter driver screening, and enhanced safety features, relying on a "price-safety" heuristic, whereby greater expenditure is equated with improved security. Thus, the seemingly contradictory results are unified: The "perception-sensitive" Class 2 manages their deep-seated safety anxiety by reducing exposure (lowering frequency to minimize anxiety triggers) and purchasing reassurance (choosing premium services for psychological safety), representing rationalized action strategies under the weight of their security concerns.

*5.3.4 Analysis of Intra-Group Heterogeneity*

To further explore heterogeneity *within* each class, Mixed Logit (ML) models nested within the LC-ICLV framework were estimated, allowing random coefficients for key variables. After stepwise screening, two variables per class passed significance tests for both the mean and standard deviation of their random parameters (see **Fig. 4**), indicating meaningful taste variation among individuals within each latent class.

In Class 1, the high standard deviation for the coefficient on Previous Ride-Pooling Experience ($\mu=0.125$, $\sigma=0.35$) indicates a bimodal preference distribution within this group – strong preference versus strong aversion. Those with a strong preference, driven by the "attribute-sensitive" nature and high price sensitivity, perceive ride-pooling as cost-effective, accepting trade-offs in comfort and privacy for perceived value-for-money. Conversely, the other subgroup exhibits strong aversion to pooling, as the additional detours, delays, and potential crowding associated with shared rides undermine comfort and convenience—key considerations for this class—particularly during nighttime travel. The smaller but statistically significant standard deviation for Platform Behavior ($\mu=0.082$, $\sigma=0.075$) indicates two more subtly differentiated subgroups. The majority trusts platform safety measures (e.g., driver verification, real-time monitoring), but a minority remains skeptical, citing concerns over data privacy and the need to disclose personal information during ride-hailing. These within-group heterogeneities further refine the profile of Class 1: while sensitive to core attributes (price, time, privacy), individuals exhibit limited but significant divergence based on personal experiences or values.

In Class 2, the coefficient distribution for Possessing a Driver's License ($\mu=0.331$, $\sigma=0.154$) indicates modest bimodality, reflecting heterogeneous psychological responses among perception-sensitive users. While some derive enhanced control from driving knowledge, aligned with earlier findings on self-efficacy. Others gain limited reassurance from such knowledge. For these individuals, acute situational anxiety—especially during nighttime travel—overwhelms the compensatory effect of driving competence and diminishes the perceived protective value of licensure, particularly in lower-tier cities. Additionally, Taxi Environment Perceived Risk (μ=-0.088, σ=0.094) shows mild heterogeneity. While most respondents exhibit aversion to unsafe environments (e.g., poorly lit streets, remote pickup points), a subset responds by adopting proactive safety strategies (such as sharing trip information or maintaining contact with others) rather than avoiding nighttime ride-hailing altogether.

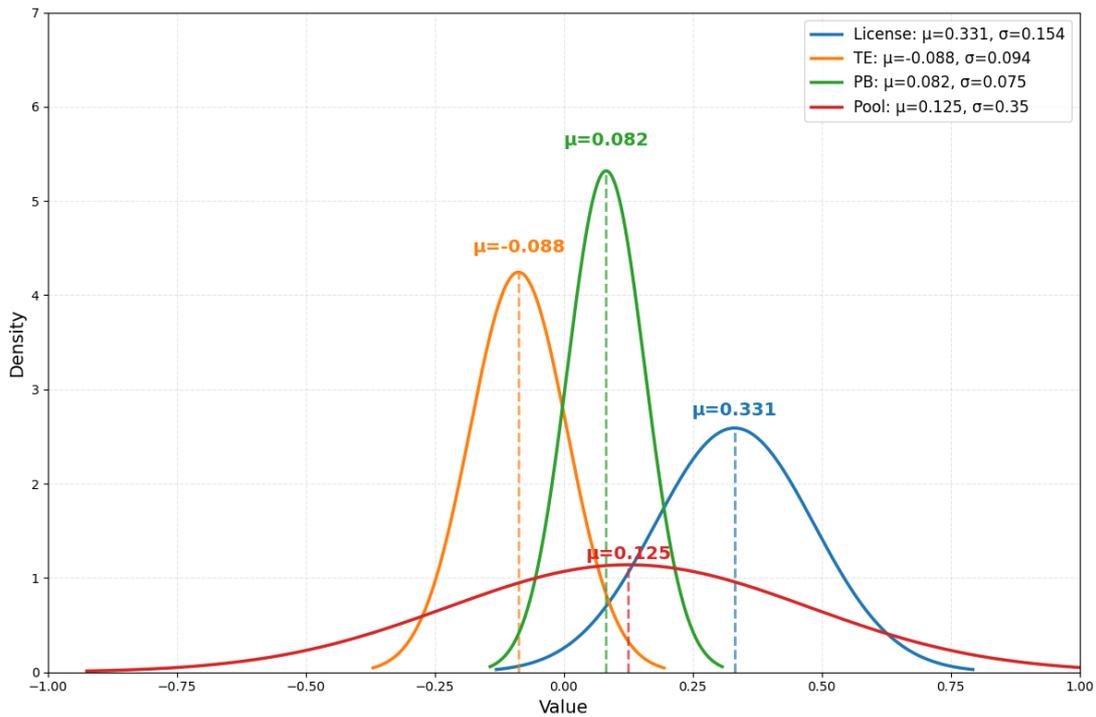

**Fig. 4.** Estimated Normal distribution of coefficients for four variables within group.

**6. DISCUSSIONS**

This study develops the LC-ICLV model to uncover heterogeneity in women's nighttime ride-hailing usage, offering a novel perspective for resolving the "convenience vs. safety" dilemma. The distinct decision-making mechanisms of Class 1 (Attribute-Sensitive Group) and Class 2 (Perception-Sensitive Group) highlight the need for a targeted intervention framework involving users, ride-hailing platforms, and regulators.

**6.1 Attribute-Sensitive Groups**

For Class 1 individuals (primarily young students and low-income women), the core challenge lies in managing the tension between affordability and safety. As 34.44% of women encountered disputes during rides (see **Fig. 2**), those low-income women, facing greater difficulties in asserting their rights, are more vulnerable to driver misconduct such as detours and arbitrary fare increases. Policy interventions should promote informed use of platform discounts during off-peak evening hours (e.g., before 9 PM) and encourage safe pickup locations, preferably well-lit, surveilled areas such as convenience stores or near police stations, avoiding secluded alleys, which pose a 70% higher risk than main roads (Peng et al., 2022). Additionally, peer-based strategies such as ride-sharing with known companions or using a "travel buddy" system have been shown to enhance perceived safety by 40% (Yang et al., 2022). Solo travelers are also encouraged to share real-time location and maintain live contact with others (family members or friends) during rides to reinforce personal security.

Ride-hailing platforms need to adjust pricing strategies and fundamental safeguards. Introducing targeted subsidies, such as "nighttime commute vouchers" or fixed-fare packages for routine trips (such as student medical visits or night-shift commutes), may improve affordability. Experience from Brazil demonstrates that cost reductions can increase women's willingness to use ride-hailing services by up to 93% (Sá & Pitombo, 2021). Ride-pooling features allowing passengers to select same-gender co-riders, alongside efforts to reduce nighttime waiting times to under 10 minutes, could further mitigate perceived risk (Guo et al., 2018). Furthermore, implementing a "safety points" incentive scheme—rewarding users for selecting well-lit pickup points or completing safety training modules with coupons or fare discounts—can reinforce protective behaviors and foster a positive safety–usage feedback loop.

Regulatory bodies should prioritize service equity by collaborating with platforms to implement targeted subsidies, such as nighttime travel vouchers for students or single mothers (Acheampong, 2021). Simultaneously, infrastructure improvements in the "last 100 meters" around schools and hospitals, such as installing solar streetlights on routes between dormitories and pickup points, can enhance perceived and actual safety. Evidence from India shows that increases in night-time streetlights have made a 5.45% rise in ride-hailing usage (Gupta et al., 2024). Regulatory oversight should also include mechanisms for rights protection, such as expedited arbitration for common grievances like detour-based fare inflation and the enforcement of joint platform liability. These measures would help ensure that economically vulnerable users are not deterred from seeking redress due to financial or procedural barriers.

**6.2 Perception-Sensitive Groups**

For Class 2 users (primarily middle-to-high-income working women), the core challenge lies in managing psychological insecurity rather than financial constraints. Despite greater purchasing power, they proactively avoid solo nighttime rides due to heightened sensitivity to perceived threats such as unexpected detours or inappropriate driver behavior. A shift from passive avoidance to proactive risk management is recommended. Women should adopt a "preventative behavior checklist", including selecting certified premium vehicles, verifying driver credentials, and activating the platform's emergency alert if the route deviates exceed 2 km. Notably, while ritualized behaviors (e.g., fake phone calls) may temporarily alleviate anxiety, they can reinforce passivity in the long term (Whitzman, 2007). Instead, instrumental strategies, such as participating in platform-sponsored self-defense programs and carrying deterrents (e.g., pepper spray or HIV post-exposure prophylaxis (PEP) kits), can enhance perceived control and promote long-term safety resilience.

Platforms should strengthen user trust through integrated technological, functional, institutional, and cultural measures. Technologically, enhancements to AI-based risk detection (e.g., fatigue driving alerts, abnormal route monitoring) should be directly linked to emergency response centers (Borres et al., 2023). Functionally, implementing a "virtual escort" feature (e.g., simulated phone calls, real-time location sharing) has shown a 68% usage rate and significant anxiety reduction (Rivera, 2007). Institutionally, enhancing transparency through regular publication of driver sexual harassment complaint resolution rates is vital; data from India indicates real-time GPS tracking can increase usage willingness in high-crime areas by 25% (Omar et al., 2022). Culturally, establishing a "Women's Safety Advisory Board" to engage female users in safety feature design ensures responsiveness to user-specific safety concerns (Zhang & Munroe, 2018).

Regulators must promote legal coordination and data sharing. Requiring ride-hailing platforms to interface with public security systems for real-time criminal background screening and to implement a 12-hour response mechanism for sexual harassment cases is critical (Ceccato & Loukaitou-Sideris, 2022). Drawing inspiration from India's Sexual Harassment Act, platforms should bear the burden of proof in sexual assault cases to enhance victim protection (National Crime Records Bureau, 2021). Furthermore, integrating transport, public security, and platform data enables dynamic optimization of police patrol routes and real-time monitoring of anomalous ride-hailing vehicle behavior, thereby improving women's safety during nighttime travel.

## 7. CONCLUSIONS

This study employs an LC-ICLV modeling framework to systematically examine heterogeneity in women's nighttime ride-hailing choices. This approach overcomes the homogeneity assumption of traditional discrete choice models by integrating latent class segmentation with an enhanced mixed logit-kernel ICLV framework. An ordered probit measurement model is used to correct for potential bias in attitudinal indicators, enabling unbiased estimation of class membership and behavioral mechanisms. The analysis uncovers two distinct segments—"attribute-sensitive" and "perception-sensitive"—and highlights the role of latent psychological constructs in shaping mode choice under conditions of perceived risk.

The findings demonstrate that women's nighttime ride-hailing behavior is shaped not by a singular risk preference or uniform safety anxiety, but by a complex heterogeneous structure influenced by socio-economic characteristics, ride-hailing service attributes, and psychological constructs. Class 1 (Attribute-Sensitive), predominantly comprised of younger women, students, low-income earners, and residents of first-tier and second-tier cities, exhibits high sensitivity of usage willingness to fare, waiting time, and platform discounts, emphasizing reliance on external service attributes over internal risk regulation. In contrast, Class 2 (Perception-Sensitive), mainly older, middle-to-high-income individuals from third-tier and fourth-tier cities, has decision-making patterns heavily dominated by perceived risk and psychological safety, manifesting as strong dependence on driver behavior, in-vehicle environment, and platform safeguards. Notably, significant micro-heterogeneity persists within both subgroups. For instance, ride-pooling experience elicits polarized reactions in Class 1, while high-frequency usage in Class 2 paradoxically intensifies avoidance behavior. This indicates that even within similar groups, individual experiences and contextual interactions can lead to variations in nighttime travel decisions. These results contribute to the behavioral literature by deepening the gendered understanding of mode choice under conditions of perceived insecurity, and offer actionable insights for multi-level interventions—including targeted fare subsidies, platform accountability, urban lighting improvements, and data-informed policing strategies.

However, this study is subject to several limitations. First, the decision outcome was modeled as a binary variable (willing/unwilling to use ride-hailing at night), overlooking women's actual range of service options, such as Hitch (private carpooling), Express (professional drivers), Ride-pooling, and Premium services. This simplified binary setting limits the ability to capture potential differences in safety perceptions and choice preferences across service types, thereby impeding the identification of specific safety concerns and their relative salience. Future research should treat service mode as a dependent variable and employ multinomial Logit models to better characterize preference heterogeneity and mode-specific behavioral responses. Second, the current choice model includes only main effects of variables, omitting potential interaction effects among explanatory variables. Incorporating interaction terms in future specifications would improve model fit and explanatory power, facilitating a more nuanced understanding of the joint influence of socio-demographic, perceptual, and contextual factors on women's nighttime mobility decisions. Addressing these limitations would yield more targeted and effective recommendations for women users, platforms, and policymakers.

**Table 9.** Measurement model estimation results.

| | | Class 1 | | | Class 2 | | |
| --- | --- | --- | --- | --- | --- | --- | --- |
| | | High-Tier City Efficient Commuters Group | | | Low-Tier City Automobile-Dependent Group | | |
| Latent Variables | Indicator | Factor Loading | Intercept | S.D. | Factor Loading | Intercept | S.D. |
| Taxi Environment | TE1 | 1(Fixed) | 0(Fixed) | 0(Fixed) | 1(Fixed) | 0(Fixed) | 0(Fixed) |
| | TE2 | **1.51\*\*\*** (40.9) | **0.071\*\*\*** (31.6) | **0.176\*\*\*** (73.1) | **1.53\*\*\*** (40.9) | **0.072\*\*\*** (21.6) | **0.185\*\*\*** (73.1) |
| | TE3 | **1.32\*\*\*** (18.3) | **0.083\*\*\*** (22.5) | **0.236\*\*\*** (78.1) | **1.61\*\*\*** (41.4) | **0.073\*\*\*** (22) | **0.176\*\*\*** (68.1) |
| | TE4 | **1.25\*\*\*** (31.4) | **0.086\*\*\*** (27) | **0.283\*\*\*** (88.3) | **1.45\*\*\*** (41.2) | **0.080\*\*\*** (24) | **0.188\*\*\*** (81.2) |
| Interior Environment | IE1 | 1(Fixed) | 0(Fixed) | 0(Fixed) | 1(Fixed) | 0(Fixed) | 0(Fixed) |
| | IE2 | **1.17\*\*\*** (21.9) | **0.121\*\*\*** (21.8) | **0.280\*\*\*** (64.3) | **1.67\*\*\*** (36.9) | **0.109\*\*\*** (31.8) | **0.181\*\*\*** (64.3) |
| | IE3 | **1.52\*\*\*** (24.5) | **0.227\*\*\*** (33.1) | **0.296\*\*\*** (79.1) | **1.74\*\*\*** (34.5) | **0.117\*\*\*** (31.4) | **0.198\*\*\*** (69.2) |
| | IE4 | **1.34\*\*\*** (24.9) | **0.127\*\*\*** (22.7) | **0.201\*\*\*** (64.5) | **1.54\*\*\*** (31.9) | **0.115\*\*\*** (32.6) | **0.200\*\*\*** (74.2) |
| Platform Behavior | PB1 | 1(Fixed) | 0(Fixed) | 0(Fixed) | 1(Fixed) | 0(Fixed) | 0(Fixed) |
| | PB2 | **1.38\*\*\*** (32.2) | **0.073\*\*\*** (32.5) | **0.291\*\*\*** (54.3) | **1.78\*\*\*** (42.7) | **0.079\*\*\*** (21.7) | **0.192\*\*\*** (64.9) |
| | PB3 | **1.64\*\*\*** (42.6) | **0.197\*\*\*** (3325) | **0.224\*\*\*** (84.6) | **1.87\*\*\*** (40.6) | **0.091\*\*\*** (23.5) | **0.204\*\*\*** (64.6) |
| | PB4 | **1.46\*\*\*** (34.3) | **0.161\*\*\*** (20.1) | **0.281\*\*\*** (75.1) | **1.76\*\*\*** (44.2) | **0.068\*\*\*** (19.5) | **0.180\*\*\*** (65.4) |

Note: \*\*\*$p < 0.01$; \*\*$p < 0.05$; \*$p < 0.1$; S.D. means Standard Deviation; The results of the t-test for each numerical value are provided in parentheses.

**Table 10.** Choice model estimation results.

| Variables | Class 1 High-Tier City Efficient Commuters Group | | Class 2 Low-Tier City Automobile-Dependent Group | |
| --- | --- | --- | --- | --- |
| | Value | t-text | Value | t-text |
| Intercept | -0.017 | -0.06 | **-0.372*** | -3.27 |
| Ride-hailing waiting time | **-1.700*** | -3.80 | **-1.230*** | -8.00 |
| Ride-hailing in-vehicle time | **-1.320*** | -5.50 | **-1.120*** | -6.20 |
| Taxi environment | -0.423 | -1.07 | **-1.110** | -1.96 |
| Platform behavior | **0.938*** | 1.74 | 0.735 | 0.91 |
| Interior environment | 0.615 | 1.27 | **0.410*** | 1.71 |
| Travel propose (Night work) | **0.506*** | 3.15 | - | - |
| Carpooling experience | **0.940*** | 1.67 | - | - |
| Encountering female driver | **-0.320**** | -1.93 | - | - |
| Maximum acceptable markup price | **0.112**** | 2.42 | - | - |
| Driving license | - | - | **0.268*** | 2.66 |
| Trip propose (Entertainment) | - | - | **0.284*** | 2.83 |
| Trip propose (Go home) | - | - | **0.237*** | 2.41 |
| Frequency (6-10 times per month) | - | - | **-0.335*** | -2.65 |
| Frequency (11-15 times per month) | - | - | **-0.298**** | -2.03 |
| Frequency (>15 times per month) | - | - | **-0.391*** | -2.53 |
| Expenditure (101-300 RMB) | - | - | **0.294**** | 2.47 |
| Expenditure (>300 RMB) | - | - | **0.503*** | 3.45 |
| Number of observations | | | 5430 | |
| Null log likelihood | | | -101179.7 | |
| Final log likelihood | | | -97395.0 | |
| Likelihood ratio test | | | 7539.3 | |
| Adjusted McFadden's rho-square | | | 0.237 | |
| AIC | | | 194858.1 | |
| BIC | | | 195082.5 | |

Note: ***$p < 0.01$; **$p < 0.05$; *$p < 0.1$;

"-" means that the variables here are not included in the model